\documentclass[manuscript,screen]{acmart}

\AtBeginDocument{%
  \providecommand\BibTeX{{%
    \normalfont B\kern-0.5em{\scshape i\kern-0.25em b}\kern-0.8em\TeX}}}

\setcopyright{acmcopyright}
\copyrightyear{2022}
\acmYear{2022}
\acmDOI{XXXXXXX.XXXXXXX}

\acmConference['XX]{Make sure to enter the correct
  conference title from your rights confirmation emai}{December 03--05,
  2022}{Woodstock, NY}
\acmPrice{15.00}
\acmISBN{978-1-4503-XXXX-X/18/06}

\begin{document}

\title{AI-Based Affective Music Generation Systems: A Review of Methods, and Challenges}

\author{Adyasha Dash}
\authornote{email: adyadash@nus.edu.sg (corresponding author), katagres@nus.edu.sg }
\email{adyadash@nus.edu.sg}
\orcid{Orcid}
\affiliation{%
  \institution{National University of Singapore}
  \streetaddress{address}
  \city{Singapore}
  \state{Singapore}
  \country{Singapore}
  \postcode{Pin}
}

\author{Kat R. Agres}
\affiliation{%
 \institution{National University of Singapore}
 \streetaddress{Singapore}
 \city{Singapore}
 \state{Singapore}
 \country{Singapore}}

\renewcommand{\shortauthors}{Dash et al.}

\begin{abstract}
 Music is a powerful medium for altering the emotional state of the listener. In recent years, with significant advancement in computing capabilities, artificial intelligence-based (AI-based) approaches have become popular for creating affective music generation (AMG) systems that are empowered with the ability to generate affective music. Entertainment, healthcare, and sensor-integrated interactive system design are a few of the areas in which AI-based affective music generation (AI-AMG) systems may have a significant impact. Given the surge of interest in this topic, this article aims to provide a comprehensive review of AI-AMG systems. The main building blocks of an AI-AMG system are discussed, and existing systems are formally categorized based on the core algorithm used for music generation. In addition, this article discusses the main musical features employed to compose affective music, along with the respective AI-based approaches used for tailoring them. Lastly, the main challenges and open questions in this field, as well as their potential solutions, are presented to guide future research. We hope that this review will be useful for readers seeking to understand the state-of-the-art in AI-AMG systems, and gain an overview of the methods used for developing them, thereby helping them explore this field in the future.
\end{abstract}



\maketitle

\section{Introduction}
Music can have a remarkable impact on listeners' emotional states \cite{juslin2011handbook}. Indeed, music is often used as a powerful medium for inducing and mediating the mood and emotional state of the listener \cite{agres2021music, clements2004use}. Meanwhile, with the advent of a new era in computing, researchers are gaining interest in designing artificial intelligence (AI)-driven music generation systems \cite{briot2020deep, herremans2017functional}. A set of these computational music composition systems focuses on creating affective music, and such systems may be referred to as ``AI-based affective music generation (AI-AMG) systems”. AI-AMG systems often have certain benefits compared to human-created music, such as the ability to skirt copyright issues, the computational means of blending genres/musical elements in novel ways, and in the case of real-time music generation systems, the ability to flexibly tailor the generated music to aspects of the environment or changes in the listeners' physical or emotional state. In addition, AI-AMG systems are potentially capable of creating an infinite number of unique affective music compositions, and composing music without any associated time constraint \cite{williams2017affective}. Due to these advantages, AI-AMG systems are now rapidly gaining the attention of researchers, as well as companies such as Google and Apple that are actively pursuing the development of creative, interactive music generation methods. 

AI-AMG systems have great potential to impact many fields including, but not limited to healthcare, co-creativity, and entertainment (gaming). By exploiting the power of affective music to induce/mediate/enhance different psychophysiological states in the listener, AI-AMG systems have been deployed in different sectors of healthcare \cite{agres2021music}. Affective music is often useful to improve the mood state of patients suffering from anxiety \cite{elliott2011relaxing} and depression\cite {stewart2019music}, while also promoting self-expression \cite{clements2004use}. Affective music can also be effective during rehabilitation, to promote physical activity or rhythmic entrainment to the beat of the music. Furthermore, emotional music can be useful in uplifting patients' mood state, thereby promoting better adherence to their prescribed rehabilitation exercises \cite{fujioka2012changes}. Thus, music-based mood mediation techniques can be applied to patients suffering from neurological disorders such as stroke \cite{dash2019design} while improving their participation in rehabilitation therapy. Another important area of application for AI-based affective music generation is co-creativity. Computational co-creativity in affective music composition refers to the collaborative composition or improvisation of music by humans and computers. AI-based algorithms can support aspects of the creative process, such as mechanizing part of the music composition (e.g., the accompaniment), thereby sharing some of the creative burden with musicians\cite{kantosalo2016modes,micchi2021keep}. Entertainment is another area in which affective music may find various applications, such as music to accompany gaming, and VR/AR-based story-telling scenarios. The use of affective music in virtual reality-based games can enhance the user's sense of immersion while facilitating their mediated presence \cite {gorini2011role}. In the case of story-telling scenarios, affective music and designated motifs can accompany certain characters and situations in the narration, to enhance the experience and better capture the attention of the listeners. Therefore, this technology-based approach to generating affective music promises to be immensely helpful in multiple fields, and can inspire composers and musicians through co-creativity.

Due to the rapidly growing interest in automatic affective music generation, it is prudent to take stock of existing systems and review the literature, both to summarize the state-of-the-art, and to help researchers working in the field gain a more thorough understanding of the most helpful techniques/methods in the area (e.g., what architectures seem most effective, what features lead to the greatest emotion induction, etc). Yet, to our knowledge, no such recent review exists. To our knowledge, only one review article \cite {williams2015investigating} has been written on this topic, which was published in 2015. Since that time, there have been huge advances in computational techniques, which have led to more advanced music generation systems. Because this review aims to assist researchers working on state-of-the-art affective music generation, this paper also focuses more on the technical aspects of AMG systems as compared with \cite {williams2015investigating}.

In this review, we 1) summarize the literature on AI-based affective music generation systems, 2) categorize these systems based on the core algorithm/method used for the music generation, 3) provide an extensive review of the different classes of AMGSs, 4) identify existing challenges in state-of-the-art AMGSs, and 4) explore potential directions for future research. 

The remainder of the article is organized as followers: Section 2 presents the method adopted for conducting this review, followed by a background of AI-based Affective Music Generation in section 3. Section 4 presents a detailed review of the literature, and Section 5 presents the important musical features and methods used to manipulate them. Challenges and future work in the field are discussed in Section 6, and the article is summarized with concluding remarks in Section 7.

\section{Methodology}

This review summarizes relevant articles from three different websites, namely, (1) Google Scholar, (2) Scopus, and (3) IEEE Xplore. These websites are mined using the search queries:  (1) (Affect OR Emotion OR Mood) AND (Synthetic OR Artificial) AND (Synthesis OR Generation) AND Music, and (2) (Affective OR Emotion OR Mood) AND (Synthetic OR Artificial) AND (Synthesis OR Generation) AND Lead sheet, (3) (Valence OR Emotion OR Affect OR Mood) AND (Generation OR Synthesis) AND (Lead sheet OR Melody OR Rhythm), and (4) (Affect OR Emotion OR Mood) AND (Generation OR Synthesis OR Composition) AND (Music AND (Lead sheet OR Score)). During the initial shortlisting process, articles from topics such as (1) Emotion-based automatic playlist generation, (2) Emotion recognition from music, and (3) Pleasant sounding artificial music synthesis, and related fields which do not address the affective component of music generation, were excluded. The remaining relevant articles that include the design of a computational affective music generation system/algorithm were shortlisted after examining the abstract. Furthermore, we looked into the references of the relevant articles in our initial search and included any additional articles published between 1990 and 2022 that met our criteria. This resulted in 56 articles in total. These 56 articles were then critically reviewed, and a detailed comparison of the articles is presented in this manuscript. Out of these 56 articles, 31 articles were published after 2015. In the next section, we present a brief background of affective music generation systems.

\section{Background: AI-based Affective Music Generation}

AI-based affective music generation (AI-AMG) is an interdisciplinary field that requires knowledge of artificial intelligence, music theory, and/or principles of music composition, as well as the fundamentals of affective science. Figure 1 illustrates the main fields that interact to create AI-based affective music, i.e., computationally-generated music that is meant to produce some perceived or induced emotion in the listener. Knowledge of artificial intelligence (AI) is important for designing an algorithm to generate automatic music. In order to compose high-quality, real-sounding music, the algorithm requires knowledge of musical rules/structure that comes from music theory and composition. In addition, concepts from affective science are useful for understanding the emotional expression of music, and how listeners respond emotionally to music. In short, AI-AMG requires interdisciplinary expertise that spans several fields, such as computing, music theory and composition, cognitive science/psychology, mathematical modeling, signal processing, and sometimes even areas such as physiology and neuroscience. This review covers the novel contributions from the main fields that contribute to the core design elements/components of an AI-AMG system, and provides a comprehensive comparison between the existing systems that are designed to generate AI-based affective music. In the following section, we present a general overview of AI-based affective music generation systems and provide a description of the major components of the system. 
\begin{figure}[!t]
\centering
    \includegraphics[width=2.5in]{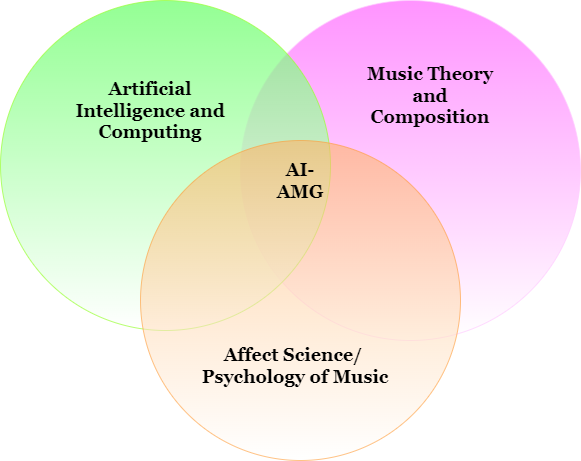}
    \caption{Major underlying fields of AI-based Affective Music Generation}
\end{figure}

AI-AMG systems usually have three major components, namely: (1) Target Emotion Identification (TEI), (2) Affective Music Generation (AMG), and (3) Emotion Evaluation (EE). Figure 2 shows a block schematic diagram of an AI-AMG system and the interaction of its components. The Target Emotion Identification (TEI) component takes input from the user or input device and maps it to the emotion domain, in a representation usable by the system. The AMG component then uses the emotion information provided by the TEI component and composes affective music accordingly. Lastly, the affective musical outputs (e.g., pieces or excerpts of music) are evaluated by the EE component (which can involve both computational and human evaluation) to examine their emotional expressiveness. A detailed description of each of these components is presented in the following subsections.

\subsection{Target Emotion Identification}

The aim of this component is to identify or select the target emotion for the affective music generation, i.e., the emotion(s) that the music aims to express or induce in the listener. This component takes the input from the user/input device in the form of text, image, video, sensor data, score, lead sheet, user input, or pre-determined target emotions (as indicated via an emotion map or discrete emotions) and provides the target emotion(s) as output, which is then used to drive the affective music generation.
The output data from the TEI component can either be presented as an element of a discrete emotion set or as a point on an emotion map (2D or 3D plane). The popular choices for discrete emotion sets include [happy and sad], and [happy, sad, calm, and angry]. These discrete emotion sets include emotions that are very distinct from one another. In the case of an emotion map, most often a circumplex model of emotion is used, such as that proposed by James Russell \cite{russell1980circumplex}, to represent the target emotion(s) on a valence-arousal plane.

The choice of input and output data for the TEI component depends on the application for which the affective music is being generated. For example, in order to generate a background music soundtrack for a gaming environment, the corresponding TEI component takes the video playback of the game as input. On the other hand, for some systems, the TEI component can take direct emotions or texts/emoji as input, depending on the application. The type of output emotion also depends on the application. For instance, in an application such as the generation of affective music for treating depressed individuals, one may wish to include a discrete set of target output emotions such as calm and happy in the TEI component.

The TEI component often uses different types of feature extraction/classification algorithms to map the input data (text, video, etc.) to the space of emotions. Depending on the application, different algorithms are used to map the video or image files onto an emotion space. For instance, a convolutional neural network architecture has been used to map input images onto a discrete emotion space (with 7 emotions) based on the emotional content in the image \cite{madhok2018sentimozart}. In some cases, the input to the TEI component is one or more emotions that come directly from the designer/user and require no algorithmic processing. In this case, the TEI component acts as a buffer and directly feeds the input emotions to the Affective Music Generation component, which synthesizes the affective music based on this information.

\subsection{Affective Music Generation: Approaches and Techniques}

The aim of the AMG component is to compose affective music that can express or induce the target emotion(s). The input to this component is the target emotion, as described above, and the output is a piece, excerpt, or continuous stream of music intended to express the target emotion. 
In the AMG component, music is usually represented as a sequence of musical events, where each event is a combination of different musical features (tempo, note duration, chord sequence, melody, harmony, timbre, etc) at different levels of hierarchy \cite {mcfee2017evaluating}. Briefly, a complex musical structure consists of low-level musical features, such as notes and chords, which are the basic building blocks for relatively high-level features, such as motives and phrases, and there exists a certain type of inter-dependency between these features in music \cite {mcfee2017evaluating}. That is, given the target emotion(s) as input, the AMG component employs particular algorithms/methods to manipulate the features and their interaction at different hierarchical levels in order to compose affective music. 

Various music generation algorithms and approaches may be used to compose the affective music, and AMGSs can be classified according to four broad types of algorithm/methods, namely, (1) rule-based methods, (2)  data-driven methods, (3) optimization methods, and (4) hybrid methods. We briefly describe each approach below. This classification is pictorially presented in Figure 3. 

\subsubsection{Rule-based methods}

Rule-based methods define musical rules that capture the relationships between musical features used for the music composition. These musical rules are represented in the AMG component either in the form of mathematical equations, or logical statements/heuristic rules (if-else statements). The AMG component uses a set of rules to compose affective music that is meant to convey or induce the target emotion(s). 

\begin{figure}[!t]
\centering
    \includegraphics[width=3 in, height = 1 in]{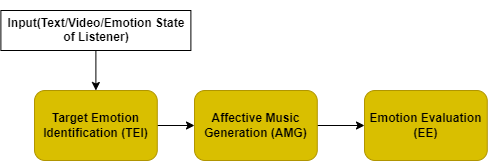}
    \caption{Components of AI-AMG system}
\end{figure}

Depending on the relationship between the musical features and the emotion, the rules can be categorized into two different categories, (1) feature-to-emotion rules, and (2) feature-to-feature rules. The feature-to-emotion category of rules depicts the relationship between the musical features and emotion dimensions (in an emotion map). For example, Wallis et al., \cite{wallis2011rule, wallis2008computer} have represented the relationship between the tempo, articulation, and roughness with different levels of arousal (an emotion-dimension). More recently, researchers have used this information to construct a more general representation of these rules in the form of parametric equations where the musical features are parameterized by levels of valence and arousal (emotion dimensions) \cite{ehrlich2019closed}. On the other hand, feature-to-feature rules depict the relationships between the features at different levels of the musical hierarchy. One such example is the ``rule for selection of a chord from a chord set to fit a given melody''. These rules can be encoded in the AI-AMG system in the form of a heuristic rule, conditional statements, or mathematical equations. A number of these rules are grouped together to form the knowledge-based of the AMG component, which then makes use of this knowledge to generate affective music. These rule-based approaches governed by a set of mathematical equations are efficient in terms of time and computation for generating music.

\begin{figure}[!t]
\centering
    \includegraphics[width=3 in, height = 2 in]{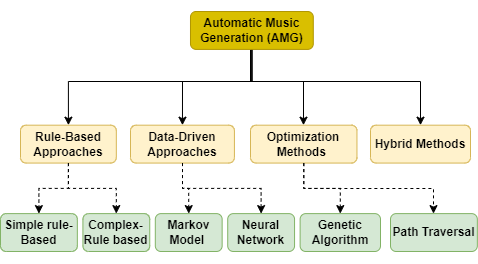}
    \caption{Classification of AMG component based on the type of algorithm/method used in music generation 
}
\end{figure}

\subsubsection{Data-driven methods}

Data-driven methods used in the AMG component capture and leverage the patters present in data to learn the innate structural information in music. Structural information is learnt during the training stage, and stored in the form of model parameters. Subsequently, the model parameters are used to generate artificial music during the generation process. This training and subsequent generation process can be conditioned on the target emotion to compose affective music. Depending on the structure of the model, data-driven methods are of two types: (1) Markov model-based approaches, and (2) Neural network-based approaches. In Markov model-based approaches, the model learns the syntactical/statistical information related to the occurrence of musical features, and subsequently, the model is used to predict the feature values that fit the musical context. Researchers have used this method to predict features such as note value, chord movement, and octave movement/register for the given musical piece \cite{ramanto2017markov}. On the other hand, neural network-based approaches use popular deep neural network architectures for learning inherent dependencies between the features of music. Later, in the generation phase, these models are used to efficiently generate new and pleasant-sounding affective music pieces. Neural networks such as seq2seq neural network architectures, Long short-term Memory (LSTM), Variational Autoencoder (VAE), and other well-known architectures are also used for composing affective music. Despite the need to train the model on a large, and often labeled dataset, the unified architecture, easy availability, easy operation, and high accuracy are some of the qualities that have made these data-driven architectures a popular choice for affective music composition.

\subsubsection{Optimization methods}

A number of optimization methods have also been successfully deployed in the automatic composing of affective music. Broadly, these optimization methods can be categorized into two categories, namely, (1) genetic algorithm, and (2) tree/graph traversal optimization methods.  

The nature-inspired optimization algorithm called genetic algorithm is used in the AMG component for generating affective music. The algorithm can be used to optimize a cost function in both constrained and non-constrained environments. The affective music synthesis procedure can be modelled as a multi-objective constrained optimization problem where the musical rules are stated as constraints. The genetic algorithm-based method can be used to find the best fit value for a music feature without violating the musical rules (stated as constraints). For example, Scire et al. used genetic algorithm for selecting the melody for a given chord progression while not violating music theoretic rules \cite{scirea2017affective}.

Other optimization techniques based on path traversal, namely, (a) tree traversal, and (b) graph traversal methods, have also been used in AI-AMG systems. These methods select the path in the graph/tree network that optimizes a given cost function. The affective music generation procedure can use a graph/tree network in which the nodes of the network depict different states of musical features, and the links between these states are weighted based on their relatedness/closeness for a given context/emotion. For example, Kou et al. \cite{kuo2015development} designed a tree network for chord selection where each node of the network is represented by a chord name, and the path length between each linked pair of chords is weighted by the valence value of the emotion. The cost function aims to optimize the total path length by selecting the closest next chord in a sequence of chords given a valence/emotion value. Similarly, graph traversal optimization methods have also been used to select chords for composing affective music \cite{scirea2017affective}.

\subsubsection{Hybrid methods}

In addition to using the aforesaid methods separately, researchers have also used a combination of these methods to compose affective music. In this case, the AMG component uses different methods (traversal methods, genetic algorithm, rules, and neural networks) to predict/select different musical features for a given emotion. Subsequently, these features are collectively presented to generate the composition. For instance, \cite{kuo2015development}, used the combination of a graph-based traversal method, genetic algorithm, and a rule-based technique for selecting the next chord in a sequence of chords, generating a melody that fits the chord progression and features such as tempo and timbre, for a given emotion/affective state.

The above AI-AMG methods are used for composing music that can express the target emotion. Subsequently, the artificially generated musical pieces are evaluated on their efficacy in expressing the given emotion. This final step is overseen by the emotion evaluation component.

\subsection{Emotion Evaluation}

The aim of this component is to evaluate the efficacy of artificially composed music to express the target emotion. Different approaches have been used by researchers for evaluating the emotional content in artificially generated music \cite{agres2016evaluation}, which can be broadly classified into two categories, namely, (1) algorithm-based assessment, and (2) human study-based assessment (depicted in Figure 4). Algorithm-based assessment methods use an analytical approach for comparing certain properties of the generated affective music with a sample template to generate a measurement index; based on the index value, the efficacy of the system in generating affect-specific music is determined. In this regard, an algorithm that computes an index by comparing the positioning of chords in a music composition relative to their position in the ``emotional circle of fifths" \cite{sergio2021scene2wav} (arrangement of chords according to their emotional expression), has been used to estimate the emotional content of the generated music. Although algorithm-based evaluation methods are easy to implement and less time-consuming than human-based assessment, such methods do not directly capture human emotional responses, and often lack the ability to measure how ``well-structured" and natural the music sounds, as well as the music's creativity (the system's ability to convey emotional content in a creative, and not overly repetitive, way). 

\begin{figure}[!t]
\centering
    \includegraphics[width=3 in, height = 3 in]{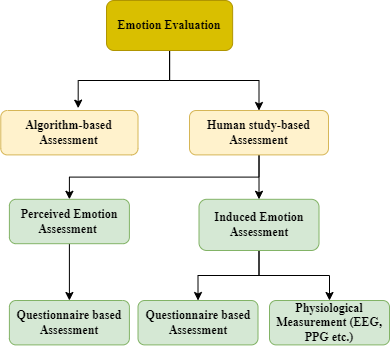}
    \caption{Classification of emotion evaluation component
}
\end{figure}

On the other hand, human study-based assessment methods \cite{ferreira2021learning, agres2016evaluation} rely on the listener’s ability to rate the emotional content of music. Emotion is a human phenomenon, and thus, using human raters provides a direct assessment of perceived/induced emotion. Even though emotion assessment by listeners can be noisy, as different individuals have different musical knowledge and preferences, such methods are often chosen by researchers. In this approach, sample music pieces from the system are played to a group of human listeners, and the listeners are asked to provide their feedback/responses \cite{ferreira2021learning} regarding the emotional content of the music. A questionnaire is used to collect subjective ratings or comments. This approach usually aims to quantify (through ratings) the emotion conveyed/expressed by the music (perceived emotion), or the emotion evoked in the listener by the music (induced emotion). Formally, as defined in \cite{kreutz2008using}, the \textit{perceived emotion ``refers to intellectual processing, such as the perception of an intended or expressed emotion"} \cite{kreutz2008using}, and \textit{``induced emotion reflects the introspective perception of psychophysiological changes, which are often associated with emotional self-regulation”} \cite{kreutz2008using}. Thus, depending on the criteria of the emotion evaluation, human study-based assessment methods can again be sub-divided into two categories, namely (1) perceived emotion assessment, and (2) induced emotion assessment \cite{wallis2011rule}. 

A conventional way of measuring perceived emotion is by collecting emotion ratings from the listeners, in terms of valence-arousal levels or discrete emotion ratings. Furthermore, induced emotion assessment methods aim to estimate the evoked emotion in the listener after hearing a certain piece of music \cite{su2018amai}. The listeners' induced/evoked emotion may be examined using a different set of questionnaires than those used for measuring perceived emotion. In addition, induced emotion can be monitored/measured by different physiological and sensor-based measurements \cite{agres2016evaluation}. For example, physiological signals such as Electroencephalogram (EEG), facial electromyogram (EMG), and Photoplethysmogram (PPG) \cite{su2018amai} may be monitored and processed for identifying and quantifying the intensity of induced emotion in listeners \cite{su2018amai, ehrlich2019closed}. Such physiology-based assessment methods are often considered to be more reliable, in the sense that these methods rely on involuntary, bias-free measures for ``directly" quantifying induced emotion. Importantly, the choice of the emotion assessment method for an AI-AMG system should be decided based on its potential application, namely whether the AI-AMG system is used to express different emotions, or is meant to induce emotions in the listener.

Even though existing AI-AMG systems are able to express the required target emotions in their generated music pieces, most researchers have not examined induced emotion in listeners, and their computational creativity has not yet matched the creative pursuit of humans -- there exists a scope for improvement. In the next section, we present a detailed review of the state-of-the-art in AI-based affective music generation systems. The papers below are briefly summarized to 1) capture which state-of-the-art methods are used, and 2) allow the reader to assess which papers are most relevant to their research (to know where to direct further reading).

\section{AI-based Affective Music Generation Systems: State of the Literature}

As discussed in the previous section, AI-AMG systems typically consist of three major components, (1) Target Emotion Identification (TEI), (2) Affective Music Generation (AMG), and (3) Emotion Evaluation (EE). Among these three components, the core functionality of the music generation is performed in the AMG component. Thus, in this section, we present a discussion of the different AI-AMGSs based on the type of algorithm/method used for music generation in the AMG component.

This article reviews 56 research papers on affective music generation systems. Of these articles, 21 articles have adopted data-driven affective music generation methods, making the data-driven method one of the most used approaches. Amongst these, 4 employ a Markov model-based approach, and the remaining 17 have deployed a neural network architecture to generate affective music. In addition to data-driven methods, rule-based methods are also popular approaches in affective music generation, with 18 articles in total. An optimization method is used in 7 articles, and a hybrid method (employing multiple of the above strategies) is used in 10 articles. Table \ref{tab:comparison} presents a comprehensive list of these articles. 

\subsection{Rule-based systems}

Rule-based AI-AMG systems make use of musical rules (i.e., the relationship between the musical features and emotion, or the relationship between musical features at different levels of hierarchy), to compose affective music. Based on the complexity of the musical rules, these systems can again be divided into two sub-categories, namely : (1) simple rule set based systems, where the rule set consists of only a small set of simple musical features such as tempo and mode (major/minor), and (2) complex rule set based systems, in which the rule set consists of rules for other musical features (pitch register, harmony, chord progression, etc.) in addition to simple musical features such as tempo and mode. Both of these approaches are used in affective music composition.

\subsubsection{Simple rule set based systems}

Some rule-based AI-AMG systems use rules that carefully select the tempo and mode of the affective composition based on target emotion information. Briefly, the tempo, measured as beats per minute (BPM), is a simple feature that has a direct and incremental relationship with the arousal component of emotion \cite{wallis2008computer}. Similarly, the mode (major/minor) of the music typically affects valence \cite{su2018amai}. These rules are widely accepted by the research community and are therefore included in most rule-based AI-AMG systems.

Of all the papers summarized in this review, only one system has used a simple rule set for generating affective content in their musical compositions. Su et al. have proposed a rule-based music generation system for improving the listener's affective state \cite{su2018amai}. The aim of the system is to compare between strategies namely, "Discharge", "Diversion" and "Discharge-to-Diversion", used for inducing relatively positive affect in a listener from his/her current emotional state. Briefly, "Discharge", "Diversion" and "Discharge-to-Diversion" are mood mediation strategies used for transferring the user from a sad/angry mood to a relatively happy mood, by playing music expressing sad/angry emotions, pleasant emotions, and sad/angry to pleasant emotions, respectively. The system uses musical rules based on two features, namely, major/minor mode and tempo, to generate musical events, and then these events are played with different combinations of instruments (piano, guitar, cello, vibraphone and kick drum). A horizontal re-sequencing method is then applied to these music events in order to generate music for the three different strategies. The aim of the music is to create a happier emotional state than the current state, and the listener's affective state is measured and quantified via their facial expression. The results indicate that the Discharge-to-Diversion strategy is better at inducing positive affect compared to the other strategies.

The simple rule-based AI-AMG system is easy to design and implement. In addition to deploying simple musical rules based on tempo and mode, complex rule-based systems make use of musical rules that control other/additional musical features based on the given emotion information. In the next section, we summarize the rule-based systems that use a complex rule set.

\subsubsection{Complex rule set based systems}

Apart from using simple musical rules for tempo and mode, AI-AMG systems have also used more complex sets of rules to carefully manipulate musical features such as harmony, melody, rhythm, etc., for creating affective composition. Such AI-AMG systems are categorized as complex rule set based systems and summarized here. These systems have been designed to operate in an open-loop mode where the target emotion information is given to the system \cite{rory2007rules}  or a closed-loop mode in which target emotion information is decoded from users' physiological state in real-time.

First, we summarize the AI-AMG systems that were designed to operate in open-loop mode. Vieillard and colleagues \cite{vieillard2008happy} developed a system that manipulates mode, dissonance/consonance, melodic pitch range, and tempo for composing affective music. Later, Wallis and colleagues \cite{wallis2011rule}  developed a rule-based system that selects pitch register, loudness, rhythmic roughness, tempo, articulation, harmonic mode, and upper extensions for composing affective music to express different levels of valence and arousal. Furthermore, researchers have also developed a set of rules named ``KTH Music Performance Rules''  for composing affective music \cite{friberg2006pdm}. Musical rules are also defined to manipulate musical features at different levels of hierarchy for composing affective music. For instance, Hoeberechts et. al presented a pipelined (sequential) architecture for synthesizing affective music by controlling the musical features at different levels of hierarchy \cite{hoeberechts2007flexible}. Here, musical rules are defined to manipulate the high-level features such as sections, blocks, and musical lines, as well as the low-level musical features, such as structure, harmonic patterns, motif patterns, modes, and meters, to synthesize happy and sad music. Recently, researchers have also tried to manipulate musical tension by changing the musical structure, i.e., varying the build-up and release of musical tension by violating musical structure (phrase violation and period violation) \cite{sun2020musical}. Such an approach of varying the tension of computerized music can also be employed for designing AI-AMGS. 

In addition, AI-AMG systems have also been designed to operate in closed-loop mode. These systems are integrated with sensor-based platforms (EEG platforms, wearable sensors) for monitoring/decoding the real-time emotional state of the user. Subsequently, based on the decoded emotional information, these systems generate adaptive emotional music. In this line, researchers have developed prototype AI-AMG systems for applications such as emotion mediation and emotion induction. Miyamoto and colleagues \cite{miyamoto2020music} developed a rule-based AI-AMG system in which the user can control the emotional music based on his/her neural activation. The emotional information is decoded from real-time EEG activation of the user and mapped onto the valence-arousal plane. Subsequently, based on this emotional information, the system synthesizes affective music by collectively changing musical features such as tempo, rhythm, loudness, mode, and pitch using a set of rules. Similarly, Ehrlich et al. \cite{ehrlich2019closed} have developed an AI-AMG system to support emotion self-regulation in listeners in real-time, by playing affective music based on the instantaneous EEG activation of the listener. The authors' AI-AMG system generates music by varying musical features including tempo, rhythm, loudness, pitch, and mode. The AI-AMG system was embedded in a closed-loop Brain-Computer Interface (BCI) system, which leveraged EEG and neurofeedback to teach the listener to mediate his/her own emotion states. The majority of the listeners were able to gain control of the music generation and self-inducing the target emotion. Kirke and colleagues \cite{kirke2013artificial} have developed an AI-AMG system that can potentially be integrated with an EEG module for user emotion mediation. In addition to EEG, wearable sensing technologies and motion/gesture sensing technologies are also employed for building a closed-loop AI-AMG system. In this endeavour, researchers \cite{wassermann2003live} have developed an affective music composition algorithm for designing an intelligent space called ``Ada". Ada tracks users' emotions by monitoring gestures (video/image,) audio (audio events), and body weight shifting profile (pressure load on the floor) which is then used to control its lighting and sound effects. The sound effect component of Ada makes use of affective music generation algorithms to generate mood-specific music by changing musical features such as scale material, timbre, tempo, sound level, articulation, and time deviations using musical rules.

In addition, some researchers have tried integrating rule-based AI-AMG systems with multimodal platforms (narrative, virtual reality, storytelling, robotic platforms) to design prototype systems for applications such as gaming and entertainment. In this endeavour, Kanno et al. \cite{kanno2015music} developed a rule-based AI-AMG system for generating affective music from input narratives. The authors first construct a musical rule base in the form of a dictionary linking a set of impression word pairs with two musical features: chord and rhythm. The impression word-pairs, such as ``Energetic - Calm”, and ``Fast - Slow”, are used to represent the two extreme impressions or emotions. A narrative is presented to the system as input and the system controls the musical features (rhythm and chord progression) using the musical rules, depending on the emotion and impression of the input narrative. A rule-based AI-AMG system has also been designed for storytelling applications \cite{da2008emotion}. In this computerized music generation system, musical features, namely, pulse salience and kinesis of rhythm, and diatonic mode (major/minor) are manipulated for encoding the emotion information in music. Similarly, the article by Nakamura and colleagues \cite{nakamura1994automatic} describes a prototype system for composing affective background music and sound effects for a given animation. The input to the system is given in the form of mood parameters, music parameters, and motion parameters for the character in the scene. The mood parameters can be glad, happy, sleepy, sad, angry, and tired, represented on a scale of 1-5. Music parameters are the musical motif, tempo and timbre, and the motion parameters comprise the actions of the character to be presented on the screen. Subsequently, these inputs are used for the selection of different musical features. For example, based on the mood information, the key of the music, the accompaniment (harmonic accompaniment-rhythmic accompaniment), and the motif information (melody of the music) are selected for generating the music. Interactive affective music generation has also been incorporated into an affect-sensitive robot \cite{huang2013study}. The platform takes emotion-related information provided by the user and further this information is used to define music parameters and the motion parameter for the robot.  Specifically, musical parameters are selected using a rule-based approach where the features such as tempo, chord (major/minor), and progression stability (chord changing rate) are determined based on the instantaneous valence and arousal values.

Most of these rule-based systems are able to reliably generate affective music pieces by encoding and then expressing the desired target emotion while using predefined musical rules. It is also observed in a rule-based system, \cite{wallis2011rule} that the perceived valence in an excerpt is dependent on the intended arousal level while perceived arousal is not affected by the valence level. In addition to expressing certain emotions through music, many of these systems can also adaptively change the emotional expression in music to transit from one emotion to another in real-time. For example, in order to change emotional expression, the music can be dynamically changed depending on the scenes of a story-telling environment, the player's situation in a game, the participant's current emotion state, for story-telling, video gaming, and mood mediation applications, respectively. Some of the articles covered in this review have also exploited the real-time transition capability of their AI-AMG systems and have integrated the system with different sensor-based platforms (EEG platforms) to work in closed-loop mode or multimodal platforms (Narrative/story-telling platforms, Virtual Reality-based platforms) for creating adaptive AI-AMG systems. Adaptive systems  (such as \cite{wassermann2003live, ehrlich2019closed}) are gaining popularity because of their potential for developing applications like interactive games, music therapy, and computer music performance. Even though these systems are capable of being deployed in myriads of applications, the design of these systems is often cited as critical. These systems often need skilled researchers with knowledge of music theory or composition to develop the rule base for the system. Furthermore, finding a convincing set of musical rules can be difficult, as the rules may differ for different genres/styles of music. Designing rules to tailor the musical feature for expressing different levels of valence has also been challenging \cite{miyamoto2020music} and needs further exploration. In spite of these limitations, rule-based systems are still a popular choice amongst researchers due to the ease in design, the requirement of fewer computational resources such as labelled datasets for training for implementation, and the ability to generate reliable, pleasant-sounding affective music.

\subsection{Data-driven systems}

One other popular way of generating affective music is by using data-driven approaches where a database of music is used to train a model, i.e., learn the parameters of the model. Subsequently, these parameters are used to generate affective music to express a certain target emotion. The data-driven model may be a Markov model where the model parameters are explicitly determined or a deep learning model where the model parameters are learned. The learning process of these model parameters can be conditioned on either discrete emotions or emotions mapped as components of valence and arousal for composing affective music. Based on the type of model used, data-driven systems are broadly classified into two categories namely, (1) Hidden Markov model-based systems, and (2) Neural network-based systems.

\subsubsection{Hidden Markov model-based systems}

AI-AMG systems have used Hidden Markov models (HMMs) for selecting chord progressions for composing an affective music \cite{morris2008exposing, simon2008mysong, soysa2010interactive}. Specifically, researchers in \cite{soysa2010interactive, simon2008mysong} have used HMM method for selecting a chord sequence to accompany a given melody for composing happy and sad music pieces. The transition probability matrices, linking chord groups, (major/minor) are used to select the next chord in a sequence for a given emotion. The validation result from the user listening study in  \cite{simon2008mysong} demonstrated that chords assigned to melodies using the HMM model receive similar subjective ratings as chords assigned manually by musicians. However, it is noteworthy here that both \cite{simon2008mysong, soysa2010interactive} the systems are designed and validated towards their potential to compose music that can convey only two discrete levels of emotion (happy and sad).

Similarly, HMM-based approaches have also been used in hybrid AI-AMG system \cite{chan2008automatic} where the next chord is selected using an HMM-based chord transition model. The detailed findings of this AI-AMG system are in section 4.5. Later, Park et al. \cite{park2010parametric} proposed a system that used hidden semi-Markov Model (HSMMs) for manipulating the emotional content in a synthesized music/song. The system controls the emotion content, represented by an Energy-Tense model (2D plane), by changing the duration and vibrato parameters of the synthesized song. The results from the listening study show that the duration parameter is more effective than the vibrato parameters in representing an emotion. 

Based on the inference drawn from the human-based assessment results/listening studies mentioned in \cite{simon2008mysong, soysa2010interactive}, HMM-based methods are efficient in encoding affective information in music, in particular for selecting chords to express the desired affect. HMM-based methods can also be a potential option for designing hybrid AI-AMG systems. However, these HMM methods are used to control only the chord progression feature, and the efficacy of this method for handling/tailoring other musical features is not fully explored. Secondly, designing an HMM model for musical features is both time-consuming and computationally expensive, as it requires a large musical dataset for constructing the transition matrix for an HMM.
In the next section, we summarize the AI-AMG systems that have used a neural network-based architecture to compose affective music.

\subsubsection{Neural network-based systems}

These AI-AMG systems have used a neural network-based architecture for generating affective music. The majority of these systems have used a (1) Long Short-Term Memory (LSTM), (2) Recurrent Neural Network (RNN), or (3) Variational Auto Encoder-Generative Adversarial Network (VAE)-GAN as the core architecture of the system.

The use of Long Short-Term Memory (LSTM) architecture has a long history in generating artificial music. The LSTM architecture has memory storing/recalling capabilities that are useful for the orderly generation of musical features in time. Thus, it has been used for generating music with a long-term coherent structure. In this endeavour,  Zhao and colleagues \cite{zhao2019emotional} have developed an AI-AMG system that uses a bi-axial LSTM architecture with a Lookback module for tackling the problem of long term-structure in music. In the bi-axial model, the time-axis and pitch-axis are used to train notes on all the pitches at each time step while the lookback module is introduced to reinforce the relationship between bars for yielding structural coherence in music.  In another system \cite{madhok2018sentimozart}, the authors have used a double-stacked LSTM architecture in conjunction with a convolutional neural network (CNN) for sentiment-specific affective music composition. In the first step, the CNN is used to extract sentiments from the input facial images, and subsequently, based on the sentiment information, the LSTM network composes affective music. In both of these systems, information related to emotion is fed to the network as a global condition in the form of emotional vectors (one-hot vectors) to compose affective music.  

Contrary to training the LSTM model with musical features, a group of AI-AMG systems have trained the LSTM model with music-coded syllables, represented as (1) music-coded-texts (musical events, a combination of musical features such as tempo, chord, etc., coded as texts) \cite{ferreira2021learning}, and (2) a musical lead-sheet \cite{makris2021generating} to address the affective music generation task as a language modelling problem. An AI-AMG system by Ferreira and colleagues \cite{ferreira2021learning} has used a  variant of the standard LSTM architecture, namely, multiplicative long short-term memory (mLSTM), trained with music-coded-texts. The trained network is used to compose affective music based on the intended sentiment, represented as levels of arousal and valence. Along these lines, Makris and colleagues \cite{makris2021generating} have developed an AI-AMG system for generating valence-conditioned lead sheets with a LSTM-based sequence-to-sequence architecture. The sequence-to-sequence framework takes the musical features as input and ``translates' high-level emotional features like valence, along with time signature, grouping, and density, into lead sheets (chords and melody). The output from the AI-AMG system is a lead sheet whose chord progression reflects the desired valence.    

In addition to LSTM architecture, RNN architecture has also been used as the core element of AI-AMG systems. The RNN can be used alone \cite{zheng2021emotionbox} or in combination with a CNN \cite{sergio2021scene2wav}  for generating emotion-specific affective music. For example, Zheng and colleagues  \cite{zheng2021emotionbox} have proposed a RNN-based AI-AMG system where musical features such as pitch histogram and note density are used for training the network. These features can be seen as representative of mode and tempo in music, and thus alteration of these features can lead to a change in emotional content in music. Exploiting this concept, the AI-AMG system has used these features for training the network instead of using a large emotionally-labelled music dataset. Later, in the generating phase, the system could generate music to express four discrete emotions, namely happy, sad, tension, and peaceful. An RNN-based AI-AMG system has also been used in video-based applications \cite{sergio2021scene2wav}, where a CNN and the RNN are used together in a sequence for generating emotion-carrying background music for emotionally annotated videos. The CNN is used to extract emotion-related visual features from the input video, and subsequently, the RNN is used to generate music based on the emotional information in visual features. 

Another popular neural network architecture used by the AI-AMG systems is the variational autoencoder and generative adversarial network (VAE-GAN) duo  \cite{huang2020emotion, qiu2019mind}. The VAE-GAN networks together form a sequence2sequence architecture connecting an encoder and a decoder/generator in series. During the training phase, the model takes music pieces, annotated with the expression of emotion, as input, and the emotion labels are fed to the network as the condition. Later, during the generation phase, the model generates affective music with the desired emotional expression. The AI-AMG systems with VAE-GAN model have been deployed to compose music with emotion tags from discrete emotion domains and different parts of the valence-arousal space.

In addition, standalone GAN, and VAE architectures have also been deployed for building AI-AMG systems.  Specifically, researchers have used GAN for generating emotion-specific melodies \cite{huang2021melody}. Similarly, the VAE architecture is also independently used in AI-AMG systems \cite{grekow2021generating, tiraboschi2021listen}. Tiraboschi and colleagues \cite{tiraboschi2021listen} have used the VAE model to design an AI-AMG system that can generate affective music in a closed-loop manner by sensing the user’s emotional state through real-time EEG measurement. Here, the user's emotions, decoded from the EEG signals, are used to drive the affective music generation process. Even simple architectures like feed-forward neural networks have been used for composing affective music \cite{williams2017affective}. For example,  the AI-AMG system in \cite{williams2017affective} uses a feedforward neural network model trained with musical features, namely, timbre, key, pitch spread, tempo, and envelope. The model is calibrated and updated by adjusting these musical features to maximize the spread of the emotional element of the generated music piece across the valence-arousal plane. The proposed system can generate affective music stimuli expressing a wide range of emotions found in the real world. Later, this AI-AMG system was integrated into a closed-loop EEG system \cite{daly2015identifying} where the user's current emotional state was used for controlling the affective music generation process.

Evaluation of the efficacy of these neural network-based systems for composing affect-specific music is conducted through user-based listening studies, and by statistical comparison with existing music \cite{makris2021generating}. The results from these studies and comparative analysis of the data confirm that the neural network-based AI-AMG systems are effective in generating affective music to express desired emotions. In fact, these systems are deployed in various fields including human-computer interaction, AI-movie creation, and emotion mediation for creating applications such as (1) design of EEG-integrated closed-loop music-based emotion mediation systems  \cite{williams2017affective}, (2) generation of music for emotional videos/games\cite{sergio2021scene2wav}, (3) creation of audio accompaniment \cite{madhok2018sentimozart, sergio2021scene2wav}, and (4) creation of X-reality \cite{tiraboschi2021listen}. However, the design of these systems requires huge computational resources and a large corpus of labelled musical data for training. Again, fine control of the emotional content in the output music from these systems is often challenging. Specific to video applications, it is also challenging for these systems to create music that can facilitate a smooth transition between different emotions. To mitigate the issue of the requirement of the large labelled dataset, researchers have tried to train the network with emotion-carrying musical features such as pitch histogram, and note density instead of training the network with a labelled musical dataset \cite{zheng2021emotionbox}. Even though the proposed method could limit the requirement related to data size to some extent, in the future researchers could aim to devise more sophisticated approaches for handling this issue along with the other challenges. Some of the potential approaches for handling these issues are also mentioned in section 6. In addition to rule-based and data-driven methods, researchers have also deployed genetic algorithm-based methods for generating affective music, which are detailed in the next section. 

\begin{table*}[]
    \centering
    \tiny
    \begin{tabular}{c|c|c|c|c|c|c} 
        \textbf{Ref} & \textbf{Generation Model} & \textbf{Emotion Model}  & \textbf{Features} & \textbf{Submethod} & \textbf{EEM} & \textbf{NDE} \\
        \hline
        \hline
        \cite{kanno2015music} & Rule-based & DE & Rhythm and Chord progression & CR & HBA & --\\
        \cite{su2018amai} & Rule-based & DE & Major/minor key, Tempo and Instrumentation & SR & -- & -- \\
        \cite{vieillard2008happy} & Rule-based & DE & Mode, Dissonance/Consonance, Melodic pitch range and Tempo & CR & HBA & 5 \\
        \cite{hoeberechts2007flexible} & Rule-based & DE & Structure, Harmonic patterns, Motif patterns, Modes, Meters & CR & -- & 2 \\
        \cite{da2008emotion} & Rule-based & DE & Pulse salience and Kinesis of rhythm and diatonic mode & CR & HBA & 4 \\
        \cite{nakamura1994automatic} & Rule-based & DE & Key, Accompaniment (harmonic, rhythmic), Motif & CR & -- & 5 \\
        \cite{robertson1998real} & Rule-based & DE & Harmony,Melody,Rhythm and Volume & CR & -- & 2\\
        \cite{miyamoto2020music} & Rule-based & V-A & Tempo, Rhythm, Loudness, Mode, and Pitch  & CR & HBA & --\\
        \cite{ehrlich2019closed} & Rule-based & V-A & Tempo, Rhythm, Loudness, Pitch, Mode & CR & HBA & -- \\
        \cite{wallis2011rule} & Rule-based & V-A & Pitch Register, Loudness, Roughness, Tempo, Articulation, Harmonic mode & CR & HBA & -- \\
        \cite{huang2013study} & Rule-based & V-A & Tempo, Chord (major or minor) and Progression stability & CR & HBA & -- \\
        \cite{friberg2006pdm} & Rule-based & V-Energy & -- & CR & -- & -- \\
        \cite{livingstone2007controlling} & Rule-based & V-A & Tempo, Rhythm, Loudness, Pitch, Mode & CR & -- & --\\
        \cite{sun2020musical} & Rule-based & -- & Musical tension  & CR & -- & --\\
        \cite{zhao2019emotional} & Data-driven & DE & -- & NN & HBA & 4\\
        \cite{madhok2018sentimozart} & Data-driven & DE & -- & NN & HBS & 7 \\
        \cite{huang2020emotion} & Data-driven & DE & -- & NN & HBA & 4 \\
        \cite{sergio2021scene2wav} & Data-driven & DE & -- & -- & HBA and ABA & 2 \\
        \cite{soysa2010interactive} & Data-driven & DE & Chord sequence & HMM & -- & 2\\
        \cite{simon2008mysong} & Data-driven & DE & Chord sequence & HMM & -- & 2 \\
        \cite{morris2008exposing} & Data-driven & DE & --  & HMM & -- & 2 \\
        \cite{ferreira2021learning} & Data-driven & V-A & -- & NN & HBA & -- \\
        \cite{williams2017affective} & Data-driven & V-A & Timbre, Key, Pitch spread, Tempo, and Envelope & NN & HBA & -- \\
        \cite{daly2015identifying} & Data-driven & V-A & – & NN & -- & -- \\
        \cite{qiu2019mind} & Data-driven & V-A & -- & NN & -- &  --\\
        \cite{makris2021generating} & Data-driven & valence & -- & NN & HBA and Statistical comparison & -- \\
        \cite{tiraboschi2021listen} & Data-driven & valence & --  & NN & -- & -- \\
        \cite{zhu2008emotional} & OM & DE & -- & GA & HBA & 2\\
        \cite{brown2012mezzo} & OM & romantic-Era Style & Harmonic tension, Formal regularities & GA & -- & -- \\
        \cite{wu2016generation} & OM & V-A & Rhythm, Volume and Pitch range & -- & HBA & -- \\
        \cite{herremans2017morpheus} & OM & -- & Musical tension & -- & -- & --  \\
        \cite{hutchings2019adaptive} & HM & DE & Harmony, Melody and Rhythm & -- & – & --  \\
         \cite{maniktala2020minuet} & HM & DE & Tempo, Chord, Note length and Octave & -- & – & 5 \\
        \cite{kuo2015development} & HM & DE & Tempo, Melody, Chord progression and  Instrument configuration & -- & -- & 12 \\
        \cite{scirea2017affective} & HM & V-A & Chord sequences,Intensity,Timbre, Rhythm and Dissonance & -- & HBA & -- \\
        \cite{scirea2017can} & HM & V-A & Chord sequences,Intensity, Timbre, Rhythm and Dissonance & -- & HBA & -- \\
        \cite{ramanto2017markov} & HM & V-A & Tempo, Pitch range, Chord dominance, Note value, Octave and Chord & -- & HBA & -- \\
        \cite{chen2013emotional} & HM & V-A &  Accompaniment, Pitch, Rhythm, Mode, Meter and Tempo & -- & HBA & --  \\
        \cite{chan2008automatic} & HM & -- &  Instrumentation, Harmonization, and Chord progression & -- & HBA & -- \\
        \cite{zheng2021emotionbox} & Data-driven & DE &  Mode and Tempo & NN & HBA & 4 \\
        \cite{de2021methodology} & OM & DE &  Scale, Tempo, Chord progression, Melody and Harmony & GA & -- & 2 \\
        \cite{kirke2013artificial} & Rule-based & V-A &  Scale, Tempo, Pitch, Loudness, Keymode & CR & – & -- \\
        \cite{xu2010emotional} & OM & DE &  Scale, Tempo, Mode & GA & HBA & 2 \\
        \cite{wassermann2003live} & Rule-based & DE &  Scale material, Timbre, Tempo, Sound level, Articulation, Time deviation & CR & HBA & 4\\
        \cite{jiang2010automated} & HM & PAD &  Pitch values, Duration between notes & -- & HBA & --\\
        \cite{huang2021melody} & Data-driven & V-A &  -- & NN & -- & --\\
        \cite{park2010parametric} & Data-driven & Energy-Tense & Duration, Vibrato & HMM & -- & -- \\
        \cite{huang2017automated} & HM & V-A & Tempo, Mode, Pitch and Chords &-- & -- & --\\
        \cite{grekow2021generating} & Data-driven & DE & -- & NN & --& 4\\
        \cite{rory2007rules} & Rule-based & -- & scale & -- & -- & --\\
        \cite{vishesh2022deeptunes} & Data-driven & -- & -- & NN & HBA & --\\
        \cite{miyamoto2022online} & Rule-based & V-A & Tempo, Rhythm, Loudness, Pitch, Chord & -- & -- & --\\
        \cite{bao2022generating} & Data-driven & DE & -- & NN & HBA & 4\\
        \cite{luo2022music} & Data-driven & V-A & -- & NN & HBA & --\\
        \cite{sulun2022symbolic} & Data-driven & V-A & -- & NN & -- & --\\
        \cite{ferreira2022controlling} & OM & V-A & -- & -- & HBA &--\\
        \hline
    \end{tabular}
    \caption{Summary Table, EEM: Emotion Evaluation method, NDE: Number of Discrete emotions targeted, DE: Discrete Emotion, V-A: Valence-Arousal Dimensions, OM: Optimization Method, HM: Hybrid Method, CR: Complex Rule-based system,SR: Simple Rule-based system, NN: Neural Network, HMM: Hidden Markov Model, GA: Genetic Algorithm, CGA: Convectional Genetic Algorithm  PAD: Pleasure-Arousal-Dominance,"--": Not available; HBA: Human-based Assessment, ABA: Algorithm-based Assessment}
    \label{tab:comparison}
\end{table*}

\subsection{Optimization method-based systems}
Based on the optimization method used in composing affective music, the systems can be categorized into two major categories, namely, genetic algorithm-based systems and tree/graph traversal optimization method-based systems. The detailed sub-categorization and the state-of-the-art are presented in the next section.

\subsubsection{Genetic algorithm-based systems}

AI-AMG systems have used genetic algorithm where the task of affective music generation systems is posed as an optimization problem that aims to find an optimal set of musical events or a set of generative musical rules (optimum weight for a set of musical rules) for composing emotion-specific music. Broadly, based on the approach used by these AI-AMG systems for updating the fitness function, the GA can be of two types. In the first type, the fitness function is a mathematical equation used for optimizing the musical rules. For instance, an AI-AMG system by \cite{de2021methodology} has used the GA method (referred to as “Conventional GA” in this article) to compose monophonic piano music with an emotion profile matching the emotion profile of the input music template. The fitness function used here computes a distance metric (mathematical equation) representing the difference in the emotional content of composed and input music pieces. The system generates music pieces by controlling the music features such as scale, tempo, chord progression, melody, and harmony. Even though the system can generate music pieces to match the emotion profile of the input music, the efficacy of the system was tested for only two discrete emotions, namely, happy and sad. 

The other type of AI-AMG systems uses a variant of a genetic algorithm, called an interactive genetic algorithm, for optimization. An interactive genetic algorithm is a form of a genetic algorithm that uses the listener's subjective evaluation information in the fitness function of the GA. Briefly, AI-AMG systems in \cite{zhu2008emotional, xu2010emotional} have used the interactive genetic algorithm (IGA) method for composing happy and sad emotional music. Here, the core algorithms use musical rules and subjective evaluation (IGA) to construct the fitness function of the GA during the training phase, and later, the updated rules are used to compose affective music. For testing the efficacy of these systems in generating affective music, authors have conducted listening studies. According to the result of these studies \cite{zhu2008emotional, xu2010emotional}, IGA method is capable of encoding the desired emotional music and also useful in composing the emotional harmonic component (consist of tempo and mode) \cite{xu2010emotional} of the music. However, the efficacy of these systems in composing affective music for multiple (more than two) emotions has not yet been tested.

Some AI-AMG systems are also designed to use composition styles (e.g., a Romantic-Era Classical style) instead of emotions for generating affective music while using genetic algorithm. For example, Brown \cite{brown2012mezzo} offers a system for composing Romantic-Era style music for computerized games using genetic algorithm. The algorithm takes the game characters and the properties of the game environment (such as props and environmental features) as the input, which are used to select the ``Leitmotivs". The system then uses the input composition (Leitmotivs) to build its composition by changing musical features such as harmonic tension and formal regularities. 

The mentioned AI-AMG systems have shown that the GA-based methods are efficient in generating affective music, and more importantly, the IGA method is a powerful way of including users’ perception-related feedback to the AI-AMG, which ultimately improves the system’s ability to compose affective music. While the use of IGA can be a better method as it requires recursive evaluation of the system by listeners, it is also time-consuming and exhausting in the training phase  \cite{tokui2000music}. However, there are only a few AI-AMG systems that have used the GA method and also the reliability of this method is tested for composing affective music only for two discrete emotions. Thus, a further investigation of the efficacy of such GA based method in generating affective music for multiple emotions is warranted.

\subsubsection{Tree/graph Traversal Optimization method-based systems}

In addition to genetic algorithm, researchers have used combinatorial optimization techniques using tree/graph traversal techniques and dynamic programming-based methods to design AI-AMG systems. In these optimization techniques, a search-based approach is used for selecting the value of a musical feature, for example, selecting the next chord value in a progression. In general, such combinatorial optimization techniques are used as a part of a hybrid AI-AMG system where the emotion-specific chord progression  \cite{scirea2017affective, scirea2016metacompose} \cite{kuo2015development} feature is chosen using a graph traversal/tree traversal method. For example, a hybrid AI-AMG system by Scirea and colleagues  \cite{scirea2017affective, scirea2016metacompose}, designed to compose affective music for an interactive gaming environment, has used the graph traversal technique for selecting an optimal chord sequence depending on the emotion presented in the game. Similarly, Kuo and co-investigators \cite{kuo2015development} have used the tree traversal method for generating emotion-specific optimal chord sequences in a hybrid AI-AMG system where the other features are selected using different methods (details of these hybrid systems and the information regarding selecting other features are summarized in section 4.5). In both of these systems, composed music is programmed to express different emotions, localized at a point on the valence-arousal plane, and the smooth transition of music between these points of emotion.

Other than combinatorial optimization techniques, dynamic programming-based optimization has also been used for designing AI-AMG system to compose user-specified emotion flow-guided  musical accompaniment \cite{wu2016generation}. Given a melody, the system generates musical accompaniment through optimal selection of chord and accompaniment pattern (rhythm, volume, and pitch range) using a dynamic programming-based constrained optimization method. Even though the system can generate affective music to convey various levels of valence and some quantized arousal levels, the effectiveness of such systems in generating music from different points on valence-arousal plane and transiting between them is yet to be verified. The variable neighborhood search-based constrained optimization method has also been used in AI-AMG system design for generating music based on a given tonal tension profile \cite{herremans2017morpheus}. This method is effective in generating music with a specified tension profile and can potentially be used for expressing emotion.

The validation results for these AI-AMG systems are promising for confirming the efficacy of optimization methods in generating affective music. In fact, the AI-AMG systems with tree/graph-based methods are mainly advantageous because they use probabilistic selection method (for musical features) that is particularly beneficial in generating non-monotonic and creative affective music. However, the major drawback in terms of designing these systems is that it requires skilled researchers with knowledge of music for designing the tree/graph structure or objective function that constitutes the core algorithm. Secondly, these tree/graph-based methods, used by these systems, can select one feature at a time, and for the selection of multiple musical features to form a composition more graph networks/other methods need to be designed and handled together. However, only chord progression is selected using tree/graph-based methods, and investigation is warranted for other musical features. In the next section, we present different AI-AMG systems that use hybrid methods used for generating affective music.

\subsection{Hybrid systems}

Hybrid methods fundamentally stand on the idea that different musical features can be tailored efficiently using different algorithms/methods for composing affective music. The hybrid AI-AMG systems use a combination of rule-based methods, data-driven methods, genetic algorithm-based methods, and graph/tree-traversal methods to compose affective music.

One group of hybrid AI-AMG systems combine HMM methods with rule-based methods \cite{maniktala2020minuet, huang2017automated, ramanto2017markov}. In these AI-AMG systems, the HMM method is used for the selection of musical chords, note length, and the octave range (the probability of playing a note from same/different octave) while musical features such as tempo, mode, and pitch range are tailored using musical rules for generating mood-specific music. For example, the hybrid AI-AMG system proposed in \cite{maniktala2020minuet} has deployed a combination of Markov model and a rule-based approach to compose affective music for a textual narrative segment. The textual narratives are first processed using sentiment analysis methods (deep learning models, support vector machine, naive Bayes classifier) to extract emotion/mood information, and subsequently, this sentiment information is used by the hybrid AI-AMG to compose sentiment-specific affective music. In another group of hybrid AI-AMG systems, instead of Markov models, different graph/tree traversal methods are used in combination with rule-based approaches \cite{scirea2017affective, scirea2017can, kuo2015development}. In these systems, the graph/tree traversal methods are opted for selecting chord progression in a polyphonic affective composition. For example, a hybrid system by Scirea and colleagues has used a graph traversal procedure to select the chord progression, and a rule-based approach to determine intensity, timbre, rhythm, and dissonance in music. Specifically, for realizing the melody component in music, the system has deployed an evolutionary algorithm (named FI-2POP) which uses a multi-constraint optimization method for minimizing the violation of musical rules during the generation of a melody. Finally, the resulting AI-AMG system can generate musical pieces for an interactive gaming platform to convey different emotions realized on the valence-arousal plane. The aforementioned hybrid AI-AMG systems are used to generate polyphonic affective music from scratch.

Instead of composing affective music from scratch, hybrid AI-AMG systems are also designed to build the affective music on a pre-composed melody\cite{chen2013emotional, chan2008automatic}. In this case, the pre-composed melody is fed to the system as input along with the emotion-related information. The system then creates the affective composition by generating the accompaniment for the given melody. For example, the hybrid AI-AMG in \cite{chan2008automatic} adapted the HMM-based method for chord selection, the nearest neighbour algorithm for selection of harmonization, and instrumentation parameters to compose affective music based on a given melody. Thus, the resulting composition is created by combining mood-specific instrumentation, melody, harmonization, and chord progression.

Hybrid AI-AMG systems have also used neural network architectures in combination with rule-based methods \cite{hutchings2019adaptive} or genetic algorithms \cite{jiang2010automated}. For instance, Hutchings and colleagues have developed an AI-AMG system to compose adaptive affective music for a multi-agent game environment. Depending on the properties of the game environment (called context), the system first retrieves a pre-composed melodic theme from a dataset \cite{hutchings2019adaptive}. This context information and the emotion information are then used to manipulate the musical features such as harmony, melody, and rhythm over the pre-composed melodic theme to create affective music. Based on the context information, an RNN is used to select the harmonic and rhythmic content for the music piece and the musical rules are used to select the melody of the composition. In a similar line, the AI-AMG developed in \cite{jiang2010automated} has combined the genetic algorithm method with neural network architecture and musical rules to compose affective music by manipulating musical features such as pitch values, duration between notes, and melody of the music.

In general, hybrid approaches are proven to be effective in generating affective music. Based on the evidence collected from human-based assessment, these systems can generate music to reliably express different emotions from discrete domains and valence-arousal plane. These AI-AMG systems are used for applications such as adaptive music generation for gaming \cite{hutchings2019adaptive}, accompaniment generation \cite{chen2013emotional}, physiology-sensitive affective music generation \cite{huang2017automated}, etc. Even though the hybrid methods are well accepted in the research community for generating affective music, there are some limitations to this approach. One of the main drawbacks of these systems is that the rationale behind the choice of different methods for manipulation/selection of different musical features is not clearly stated. This challenge is termed the problem of “hybridization” which is detailed in section 5 with some possible solutions. The second major drawback in designing hybrid systems is the need for skilled researchers with knowledge of music theory. Specifically, as a part of the hybrid system, designing graph/tree traversal architectures, HMM models and musical rules demands a deeper understanding of music theory to construct the models.
In the next section, we provide a comprehensive overview of the algorithms used to create affective music, focusing on the musical features they employ to control and generate the affective music.

\section{Guideline(suggestions) for (on) designing affective music generation systems}

In this section, we will briefly summarize the most reliable features in affective music generation that are used for encoding affect information in music and also discuss the procedures that can be potentially employed for tailoring them. The general procedure for creating an AI-AMG system includes four steps:(1) Establish research goals and functional application of the system, (2) Based on 1, and the capabilities of the research team, select appropriate methods/AI-AMG system architecture, (3) Implement the AI-AMG system, which can manipulate musical features in order to achieve the goals of 1, (4) Evaluate the efficacy of the system. This review has focused on presenting the various architectures used for affective music generation; we, therefore, focus on steps (3) and (4) in the remainder of this section.

\subsection{Selecting features for affective music generation}

Table \ref{tab:feature} summarizes the important musical features of affective music generation and their dependency on different aspects of emotion (valence and arousal). Various musical features have an impact on emotions. Different algorithms can harness (all of) these features, but some algorithms may be able to more effectively control some features than others, and thus are used by researchers more frequently. (Note: Here, for presenting the features and their dependence on emotion dimensions in a concise manner, we consider each feature independently and present their dependence on valence, arousal, or both (i.e., ``valence" or ``arousal", ``valence and arousal"). The relationship between individual features and discrete emotions is not included in the table because the information related to discrete emotions is often encoded into an affective composition by manipulating multiple features of the composition collectively (not individually).

First, we provide details of some important features that are tailored by the majority of researchers to embed emotion in music. One often manipulated feature is tempo (beats per minute). Rule-based approaches are effective in harnessing the ability of tempo to convey emotional information \cite{wallis2008computer, ehrlich2019closed, vieillard2008happy}. The tempo of a composition \cite{wallis2008computer, ehrlich2019closed} is normally controlled based on the arousal value of the target emotion where the mapping between tempo and arousal is governed by an algebraic equation. Nevertheless, methods such as neural networks \cite{williams2017affective} and interactive genetic algorithms \cite{xu2010emotional} have also been effectively used for manipulating this feature. Moreover, in all of these systems, the rule of thumb for manipulating tempo is that faster tempos are associated with higher arousal. 

Another very important feature in affective music composition is the mode of the composition, e.g., the scale (typically major/minor) in which the music is composed, which is linked to emotional valence. A rule-based approach can be used for selecting the mode of an affective composition \cite{su2018amai, zheng2021emotionbox}, where positive valence is associated with the selection of major chords, and negative valence is associated with minor chords. A combination of mode and tempo has also been used by researchers (e.g., \cite{vieillard2008happy, williams2017affective}) for encoding different combinations of valence and arousal in an AI-AMG system.
Apart from selecting the scale/mode, the selection of a sequence of chords, named chord progression, is often cited as an important feature of affective composition, and the selection of the chord progression is usually based on valence \cite{kuo2015development}. Probabilistic approaches such as hidden Markov models \cite{simon2008mysong} and tree traversal/graph traversal methods \cite{scirea2017affective, kuo2015development} have shown promise in selecting the chord progression to elicit particular emotional responses. It is noteworthy that these approaches are not only effective in selecting the next chord for the composition but also in potentially bringing some amount of variation to the composition due to their probabilistic nature. 

The pitch/pitch register characteristic in music is also crucial for expressing particular emotions through music and can be defined by the selection of a certain pitch register or selection of the allowed range for the pitch registers. The pitch register characteristics can be selected based on valence \cite{wallis2008computer} or arousal \cite{wu2016generation}. The literature shows that rule-based approaches for selecting the probability of a certain pitch register depend on valence levels \cite{wallis2008computer}, although a rule-based method for defining the allowed range of pitches has also been successfully deployed based on arousal \cite{wu2016generation}. Such methods can be potentially considered in the future for selecting pitch characteristics in affective music composition.

Instrument Volume (dynamics) has also been considered as a critical parameter for affective music generation, which can carry relevant information about both valence and arousal \cite{kuo2015development}. Rule-based methods have proven to be effective in selecting the instrument volume in a composition \cite{kuo2015development}. Rhythm is another important feature that can encode and convey affect in music composition. Rhythm is defined by relative note durations and their temporal organization \cite{da2008emotion}. Rhythms in a composition may be selected based on arousal levels alone \cite{wallis2008computer}, or combinations of valence and arousal levels \cite{scirea2013mood}. In order to modulate rhythmic components, a simple rule-based approach (e.g., \cite{wallis2008computer}) such as a probabilistic note-onset rule for given arousal values is used. In addition, a complex rule-based approach for generating Euclidean rhythms has also been effectively used for composing rhythmic component in an affective composition while tailoring the rhythmic component based on both valence and arousal values \cite{scirea2017affective}. 

Above we discussed some potential features of affective music composition that could be tailored independently based on given emotion components (valence and/or arousal). However, the majority of affective composition systems are designed to compose music by simultaneously modifying multiple features for a given valence/arousal value or discrete emotion. In order to simultaneously modify multiple features, a combination of different approaches has been considered in the literature \cite{scirea2017affective, kuo2015development}. Hybrid methods may be a powerful choice, as they can provide the flexibility of tailoring different features using their most suitable algorithms, and then generating the music by combining all of these features. This approach can be pursued further in the future with the aim to find a suitable and feasible combination of features along with their tailoring algorithms for composing affective music.
Apart from hybrid methods, data-driven (specifically neural network) approaches have also proven to be efficient and effective in handling multiple features at a time to compose affective music. In particular, sequential architectures such as RNNs and seq-to-seq models have proven to be effective in generating affective compositions. Even though such architectures have been effective in generating music with the required emotion, a great deal of work remains in terms of precisely controlling the emotional content of the music generated in this process \cite{carnovalini2020computational}. This is often referred to as the ``challenge of controlling the affect" in artificial music composition (a detailed discussion is in section 6.2).
\begin{table*}[]
    \centering
    \small
    \begin{tabular}{c|c|c}
        \textbf{Feature} & \textbf{Method/algorithm most frequently adopting the feature} & \textbf{Emotion component} \\
        \hline
        \hline
        Tempo & Rule-based/Neural Network/Interactive Genetic Algorithm & Arousal \\
        Mode/Scale (major or minor) & Rule-based & Valence \\
        Chord progression/Sequence & Tree traversal/Graph traversal/ Hidden Markov model method &  Valence\\
        Instrument Volume &	Rule-based &	Valence and arousal\\
        Rhythm &	Rule-based &	Arousal/ Valence and arousal\\
        Pitch Characteristics &	Rule-based &	Valence/ Arousal\\
        \end{tabular}
    \caption{Musical features, their impact on different dimensions of emotion (valence and arousal), and the methods that most often employ each feature.}
    \label{tab:feature}
\end{table*}

In the next section, we discuss the major challenges in designing computationally creative systems, with a focus on AI-based affective music generation systems, and will also discuss possible directions for improvement.

\section{Discussion}

\subsection{Bottlenecks in state-of-the-art systems}

In the previous section, we have presented detailed state-of-the-art methods used for affective music generation and summarized the AI-AMG systems where these methods are used. A comparison is presented in Table \ref{tab:comparison}. These AI-AMG systems have represented the emotions on a valence-arousal plane or as discrete emotion sets. We care to mention here that there is a lack of comparability possible across these systems because of the disparity in the selection of the (1) number of discrete emotions, and (2) listener's group in the validation study.

The disparity in the choice of the number of discrete emotions can be termed as ``variability in the number of discrete emotions". For example, some of the AI-AMG have targeted four discrete emotions while some of the AI-AMG systems have tested only two discrete emotions (happy and sad). Systems capable of expressing more discrete emotions are better capable of precisely tailoring the musical features and can be seen as superior to their counterparts. On the other hand, for the emotion evaluation process, the majority of these AI-AMG systems have relied on human study-based assessment methods. While human study-based assessments are advantageous due to their inherent ability to capture human perception (on emotion and creativity in music) during the evaluation process, the criteria for the selection of the listener's group were not carefully chosen. The listeners' groups, used in the validation studies, have shown high variability in the number of listeners, demographics of listeners (age, gender), and prior musical training. This variation in listeners' group selection can be termed as ``variability in listener group". The ``variability in listener group" might have a significant confounding effect on the validation of AI-AMG and in the future, this music listening study group needs to be standardized. The comparison presented in Table \ref{tab:comparison} is susceptible to the effect of these variabilities (selection of discrete emotion set and listener group for validation).

In the next section, we discuss the major challenges and open research questions in designing a reliable AI-AMG system that is capable of generating real-sounding, affective music.  

\subsection{Challenges}

In this section, we discuss the open challenges in designing AI-based automatic, affective music generation systems. The pictorial representation of the challenges is given in Figure 5. We hope that this discussion will help readers who aim to design AI-based affective music generation systems capable of producing real-sounding, affective music using computational creativity \footnote {Computational Creativity is a multidisciplinary field that tries to obtain creative behaviors from computers {\cite{colton2012computational, carnovalini2020computational}}}. 

One of the major challenges in creating AI-AMG systems is ``Control”, which in this case refers to allowing the user to specify the desired emotional content of the generated music (and the ability of the system to precisely control the musical features so that the resulting music exhibits the desired affective information \cite{carnovalini2020computational}). Unlike human composers who are privileged to adapt musical patterns/ideas \cite{carnovalini2020computational}, AI-AMG systems only rely on the learned parameters/information gained during training to compose the music (which is especially true for data-driven systems/neural network systems). Thus, for such systems, controlling the features to produce desired output often becomes difficult due to limited transparency in the input, output, and their interdependencies \cite{briot2020deep}. In order to address this issue of controllability, researchers have developed conditioning strategies by using conditional architectures. Here, the data-driven architecture/neural network model has an additional condition imposed during the training phase. For example, Markis et al. have proposed an AI-AMG system with high-level conditional information that is fed into the encoder part of a sequence-to-sequence architecture for generating valence-specific affective music in a more controllable manner (e.g., the music is generated to match a profile of valence values provided by the user) \cite{makris2021generating}. However, a challenge in this direction is that datasets labelled with reliable affective information are scarce, although some researchers are working on this limitation \cite{chua2022predicting}. Apart from using a conditional architecture, another potential approach to improve controllability is to use the fundamentals of reinforcement learning to train the neural network models \cite{briot2020deep}. Briefly, the idea is to frame the music generation process as a reinforcement learning problem, where the objective function of the model is a combination of the objective function of the recurrent network along with some user-defined constraints. Such methods have proven to be beneficial in improving the controllability of automatic music generation systems  \cite{briot2020deep}, and this approach can also be considered in the future to improve the controllability of the emotional content of the generated music.
\begin{figure}[!t]
\centering
    \includegraphics[width=3 in, height = 3 in]{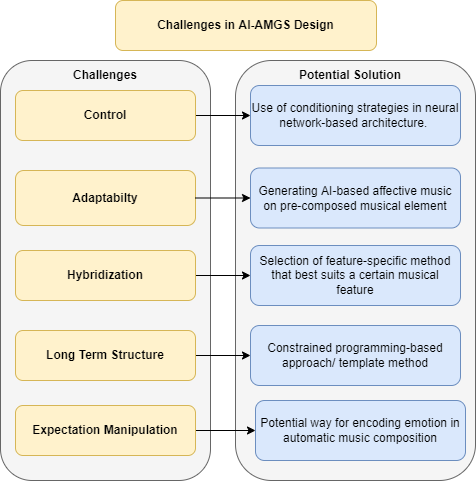}
    \caption{Challenges in designing AI-AMGS }
\end{figure}

The second major challenge is ``Adaptability”, also called ``Narrative Adaptability,” which refers to the capability of the AI-AMG system to generate a coherent piece of music that can adaptively change based on the given narrative/sequence (emotional requirement of the narration) of a story. One of the potential applications for AI-AMG systems is to compose music that is able to match the dynamically changing events in the narrative/story-telling by adaptively controlling the emotion expressed in the generated music. It is often difficult, however, for the AI-AMG system to control/tailor its musical features to reliably convey the transitions between different events in the narrative. It is worth mentioning here that this task of generating adaptive music for narratives cannot be accomplished using human-composed music alone, and thus can be seen as an interesting challenge for AI-AMG systems \cite{carnovalini2020computational}. One potential solution to this problem is co-creativity (using a combination of human- and machine-generated music). In this regard, an AI-AMG system named ``Mezzo” \cite{brown2012mezzo} developed by Daniel Brown composes romantic era-style music for video gaming. Here, a theme (Leitmotiv), related to an in-game scenario that involves particular characters and situations, is composed by humans, and these themes are given as input to the system. Later, when the event is encountered in the game, the AI-AMG system composes music by blending the human-composed theme with the computationally-generated music, in order to express the appropriate in-game situation and emotion. Another possible way of improving adaptability is to first understand the interplay between features with respect to different emotions, study how each specific feature can be independently tailored to convey a particular emotion, and then adaptively change a single (most reliable) feature based on the narration while keeping others constant. Such an approach, also proposed by \cite{carnovalini2020computational}, can be beneficial for allowing the AI-AMG systems to carefully adapt their emotional expression while answering some fundamental questions in this field.

Some other pertinent challenges in designing AI-based affective music generation systems are hybridization, long-term structure, and manipulation of the listener’s expectations. Hybridization refers to combining more than one technique for music generation in a single AI-AMG system. Even though researchers have deployed many different combinations of techniques (details in section 4.5) for designing AI-AMG systems, the rationale behind the choice of these techniques and combining them is often unclear. In this regard, a more systematic approach can be helpful, where the selection of different techniques for different musical features can be optimized based on their efficacy in generating affective music. The problem of long-term structure refers to the difficulty in generating longer excerpts that have an adequate amount of musical structure, such as repetitive patterns, in order to sound musically-coherent across the composition. This issue of long-term structure is a general problem for automatic music generation system and is not only limited to affective automatic music generation systems. In order to tackle this problem, researchers, for example have developed a constrained programming-based approach for preserving the tension profile of a given music piece to generate a new music piece. This method is efficient in generating automatic music with long-term structure \cite{herremans2017morpheus}, and such an approach can also be potentially translated and implemented in AI-based affective music generation systems. 

In addition, the manipulation of musical expectation in a composition can be seen as a viable method for eliciting different emotions in a listener. More precisely, the process of emotion induction via music can be seen to be fundamentally governed by a reward mechanism in the brain that responds to the realization and violation of musical expectations over time \cite{cheung2019uncertainty}. Therefore, another possible way of encoding emotion in music would be by algorithmically manipulating musical expectations in the piece. This could create a new avenue of research in the field of AI-based affective music generation systems. Thus, finding a feasible algorithmic approach for explicitly manipulating the expectation in music can be posed as a future research question. 

\section{Conclusion}

In this article, we have presented a comprehensive review of AI-based affective music generation systems. The main components of AI-AMG systems were detailed, and different approaches used for designing these components were presented. Subsequently, we reviewed and summarized the state-of-the-art methods that have been deployed to develop reliable AI-AMG systems. For the reader’s interest, we presented a set of important features to include in such systems, their relationship with arousal and valence, as well as previously used methods for controlling them. Finally, we have summarized the challenges and open questions in the field of developing reliable affective automatic music generation systems. We hope that this review will be useful for readers who seek to understand the different AI-AMG systems that have been developed and acquire an overview of the methods used for developing them, thereby helping future exploration of this field. We hope that this review is also helpful to researchers entering this field as they frame their research questions for developing new AI-AMG systems.

\bibliographystyle{ACM-Reference-Format}
\bibliography{refs}










\end{document}